\documentclass{article}

\usepackage[utf8]{inputenc}
\usepackage{times}
\usepackage{graphicx} %
\usepackage{subfigure}
\usepackage{natbib}
\usepackage{hyperref}
\usepackage{kitcolors}
\usepackage{bm}
\usepackage{balance}

\usepackage{siunitx}
\usepackage{amsmath}

\usepackage{algorithm}
\usepackage{algorithmic}
\usepackage[acronym]{glossaries}
\glsdisablehyper
\usepackage{standalone}
\usepackage{tikz}
\usetikzlibrary{shapes.multipart, positioning, matrix, arrows.meta}
\usepackage{pgfplots}
\pgfplotsset{compat=1.17}
\usepackage{csvsimple}
\usepackage{booktabs}

\newacronym{gcs}{GCS}{geometric constellation shaping}
\newacronym{pcs}{PCS}{probabilistic constellation shaping}
\newacronym{ccdm}{CCDM}{constant composition distribution matching}
\newacronym{ess}{ESS}{enumerative sphere shaping}
\newacronym{oess}{OESS}{optimum enumerative sphere shaping}
\newacronym{awgn}{AWGN}{additive white Gaussian noise}
\newacronym{fec}{FEC}{forward error correction}
\newacronym{ask}{ASK}{amplitude shift keying}
\newacronym{qam}{QAM}{quadrature amplitude modulation}
\newacronym{lut}{LUT}{lookup table}
\newacronym{pas}{PAS}{probabilistic amplitude shaping}
\newacronym{air}{AIR}{achievable information rate}

\usepackage[accepted]{grcon}
\graphicspath{{images/}}

\pgfplotsset{
  discard if/.style 2 args={
    x filter/.append code={
      \edef\tempa{\thisrow{#1}}
      \edef\tempb{#2}
      \ifx\tempa\tempb
        
      \fi
    }
  },
  discard if not/.style 2 args={
    x filter/.append code={
      \edef\tempa{\thisrow{#1}}
      \edef\tempb{#2}
      \ifx\tempa\tempb
      \else
        
      \fi
    }
  }
}
\newcommand\extrafootertext[1]{%
  \bgroup
  \renewcommand\thefootnote{\fnsymbol{footnote}}%
  \renewcommand\thempfootnote{\fnsymbol{mpfootnote}}%
  \footnotetext[0]{#1}%
  \egroup
}

\grcontitlerunning{Introducing RSESS: An Open Source Enumerative Sphere Shaping Implementation Coded in Rust}

\begin{document}

\twocolumn[
\grcontitle{Introducing RSESS: An Open Source Enumerative Sphere Shaping Implementation Coded in Rust}

\grconauthor{Frederik Ritter}{uoogk@student.kit.edu}
\grconauthor{Andrej Rode}{rode@kit.edu}
\grconauthor{Laurent Schmalen}{laurent.schmalen@kit.edu}
\grconaddress{Karlsruhe  Institute of Technology (KIT), Communications Engineering Lab (CEL), Hertzstr. 16, 76187 Karlsruhe}

\grconkeywords{software radio, gnu radio, dsp, GRCON}
]

\vskip 0.3in

\begin{abstract}
In this work, we present an open source implementation of the \gls{ess} algorithm used for \gls{pcs}.
\Gls{pcs} aims at closing the shaping gap caused by using uniformly distributed modulation symbols in channels for which information theory shows non-uniformly distributed signaling to be optimal.
\gls{ess} is one such \gls{pcs} algorithm that sets itself apart as it operates on a trellis representation of a subset of the possible symbol sequences. \gls{ess} leads to an empirical distribution of the symbols that closely approximates the optimal distribution for the \gls{awgn} channel.
We provide an open source implementation of this algorithm in the compiled language Rust, as well as Python bindings with which our Rust code can be called in a regular Python script.
We also compare simulation results on the AWGN channel using our implementation with previous works on this topic.
\end{abstract}

\section{Introduction} %
The capacity of a channel is defined as the maximum code rate with which reliable transmission (i.e. with vanishing error probability) is possible.
On an \gls{awgn} channel, the capacity can be achieved with a continuous and normally distributed channel input~\cite{shannonMathematicalTheoryCommunication1948, forneyMultidimensionalConstellationsIntroduction1989}.
Though this would be optimal for a continuous channel input, a discrete set of channel input symbols is used in practical communication systems.
Furthermore, many communication systems employ a set of uniformly distributed, discrete symbols as channel input.
As a result, the channel capacity can not be achieved.
The gap to capacity caused by using a suboptimal channel input is called shaping gap and amounts to \SI{0.255}{bit/channel\ use}~\cite{Gultekin2020, Forney1984}.
In terms of signal-to-noise ratio (SNR), this corresponds to a loss of \SI{1.53}{dB} in energy efficiency. 

Two major approaches to reduce the shaping gap are known in literature~\cite{sunApproachingCapacityEquiprobable1993,kschischangOptimalNonuniformSignaling1993}:
\Gls{gcs} and \gls{pcs}.
While \gls{gcs} changes the constellation symbols and induces changes to most parts and algorithms in the communication system, \gls{pcs} alters the probability of occurrence of constellation symbols placed on a rectangular, evenly spaced grid.

One difficulty with probabilistic constellation shaping is the integration with \gls{fec}.
The de-mapping of received symbol sequences back to bit strings is sensitive to wrongly detected symbols and does not easily allow the use of soft information.
In \cite{Boecherer2015}, the \gls{pas} architecture was introduced to mitigate this problem.
It works by shaping only the amplitude of transmit symbols and the approximately uniformly distributed parity bits are used to determine the sign. At the receiver the channel decoder can use soft information to recover the shaped bits which were used to create the amplitude sequence. Therefore the amplitude sequence can be regenerated error-free and dematched to the original bit sequence.
The \gls{pas} architecture combines the benefits of probabilistic shaping with the benefits of using \gls{fec}.
Because this is an important improvement over plain probabilistic shaping, for the remainder of this paper we will assume the use of \gls{pas}.
Therefore, further discussion will focus on mapping a sequence of bits to a sequence of amplitudes rather than to a sequence of symbols.
We have to note that this approach only works for distributions which are symmetric in their amplitudes.

\Gls{pcs} can be subdivided into direct and indirect methods.
Direct methods attempt to change the occurrence probability of the transmit symbols to a given target distribution.
A prominent example of the direct method is \gls{ccdm}.
It works on fixed-length symbol sequences by collecting into a code book only those sequences, where the relative frequency of occurrence of the symbols matches the desired probability of occurrence.
Bit strings are then unambiguously assigned to the sequences in the codebook and the corresponding sequence is sent in place of a given bit string.
The decoder in the receiver uses the same codebook, such that it can recover the original bit string from the received symbol sequence.
Using arithmetic coding~\cite{schulteConstantCompositionDistribution2016}, the mapping and de-mapping of \gls{ccdm} can be implemented efficiently.
Unfortunately, this straightforward scheme suffers from significant rate losses if the sequence length is short.
Other direct methods, like multiset-partition distribution matching~\cite{fehenbergerMultisetpartitionDistributionMatching2019}, try to alleviate this disadvantage.
This paper focuses on indirect \gls{pcs} methods, which induce a desired probability distribution through a sufficiently well-designed goal function.
In  the context of Gaussian channels, one possible goal function limits the energy of the fixed-length symbol sequences in the codebook.
This approximates a Maxwell-Boltzman distribution of the symbols for large sequence lengths, which is the optimal distribution for discrete symbols~\cite{kschischangOptimalNonuniformSignaling1993}.
By including all sequences with energy below a certain threshold,
indirect methods create the largest possible codebook for a given average energy.
As the shaping rate is proportional to the logarithm of the codebook size, they suffer the minimal rate loss achievable with a finite sequence length.
By interpreting a sequence of symbols as a multidimensional vector, the energy of the sequence becomes the vectors' square norm.
All sequences with their energy lower than the threshold would thus be contained in a multidimensional sphere.
Hence, these methods are also called sphere shaping.
There are multiple algorithms that use sphere shaping, notably: Laroia's first algorithm, shell mapping \cite{Laroia1994}, and \gls{ess} \cite{Willems1993}.

\Gls{ess} uses a trellis representation of the codebook and performs the mapping to bit sequences based on a lexicographical ordering of the symbol sequences.
This allows for a slight reduction in complexity compared to Laroia's first algorithm and a substantial reduction in complexity compared to shell mapping.
A drawback of \gls{ess} is that the lexicographical indexing leads to slightly suboptimal results if the number of sequences is limited to a power of two.
This is relevant because the number of bit sequences of a fixed length is always a power of two. \cite{Gultekin2020}

\emph{Notation}:\\
\Gls{ask} is a modulation scheme that encodes the information in multiple real symbols.
Using individual \gls{ask} constellations for the inphase and quadrature branch of an IQ modulator, \gls{qam} follows.
As we are only interested in the amplitudes for shaping, we define the set of amplitudes for an $M$-ASK system as
$$
\mathcal{A} = \{1, 3, 5, ..., M-1\}.
$$
A sequence of $N$ amplitudes is denoted by $\bm{a}^N \in \mathcal{A}^N$.
The individual amplitudes in the sequences are denoted by
$$
\bm{a}^N = (a_0 \; a_1 \; a_2 \; \cdots \; a_{N-1}).
$$
We use the squared norm of an amplitude sequence to define its energy
$$
E(\bm{a}^N) = ||\bm{a}^N||^2 = \sum_{n=0}^{N-1} \bm{a}_n^2.
$$

The remainder of this paper is organized as follows:
We first provide an overview of the \gls{ess} algorithm in Section~\ref{sec::ess}.
A discussion of the \gls{oess} algorithm, which addresses the issue of \gls{ess} being suboptimal for fixed bit length indexes, is added in Section~\ref{sec::oess}.
The introduction and evaluation of RSESS, which implements \gls{ess} and \gls{oess}, follows in Section~\ref{sec::rsess}.
Finally, the paper is summarized by Section~\ref{sec::conclusion}.

\section{Enumerative Sphere Shaping} \label{sec::ess} %

In this section, we will briefly outline the algorithms used in \gls{ess} for mapping from a bit-sequence to a symbol sequence and vice versa. We like to refer the reader to~\cite{Willems1993} for the original idea and to~\cite{Gultekin2020} for a more detailed description.
\subsection{Bit Sequence to Amplitude Sequence Mapping}

To transform a stream of uniformly distributed bits into a stream of non-uniformly distributed symbols, \gls{ess} uses a fixed-to-fixed length mapping:
A fixed-length sequence of bits is transformed into a fixed-length sequence of symbols.
The possible symbol sequences are collected into a codebook and,
as the bits are uniformly distributed, all symbol sequences in the codebook are equally likely. %
To achieve a non-uniform symbol distribution, the symbol sequences in the codebook have to be chosen carefully.
In \gls{ess}, this is achieved by constructing a codebook of all sequences with energy less than a fixed energy threshold $E_\text{max}$.
For an infinite sequence length, the symbol distribution in this codebook converges to the Maxwell-Boltzman distribution.
In addition, the average energy of the codebook is always minimal for its size, which leads to minimal rate loss.
The one-to-one mapping from bit sequences to symbol sequences is obtained by lexicographical ordering of the codebook.
Lexicographical ordering is the method of ordering words in a dictionary but applied to sequences of symbols.
A sequence $\bm{a}^N$ is said to be larger than  sequence $\bm{b}^N$ if there exists some $n$, with $1 \leq n \leq N$ such that the symbols of both sequences below index $n$ are equal $\left(a_i = b_i, i<n\right)$ and its symbol at index $n$ is larger than that of the other sequence $\left(a_n > b_n\right)$.
For example, the first sequences in the codebook for a sequence length $N = 3$ and $8$-ASK are $(1 \; 1 \; 1), \; (1 \; 1 \; 3), \; (1 \; 1 \; 5), \; (1 \; 1 \; 7), \; (1 \; 3 \; 1), \; (1 \; 3 \; 3)$ and so on.
Having defined an ordering allows indexing the sequences.
The index $i(\bm{a}^N)$ of a sequence $\bm{a}^N$ is defined as the number of sequences below it.
Thus with the example from above, we can state that $i((1 \; 1 \; 1)) = 0$, $i((1 \; 1 \; 3)) = 1$, $i((1 \; 1 \; 5)) = 2$ and so on.
Due to the mapping being invertible, we can easily define the inverse mapping $\bm{a}^N(i)$ as the sequence with index $i$.
Taking the energy threshold into account, not all possible sequences are contained in the codebook.
Table~\ref{tab::codebook-lut} shows all amplitude sequences in the codebook for $N = 4$, $E_\text{max} = 28$ and $8$-ASK.
Each index can be converted to its binary representation to obtain an invertible mapping from bit sequence to amplitude sequence.

\begin{table}[h]
	\caption{Codebook for $N = 4$ and $E_{\text{max}} = 28$ using $8$-ASK with index for each sequence according to
    \cite{Gultekin2020}.}
	\centering
	\begin{tabular}{cc cc cc}
 \\
    \toprule
	$i$ & $a^N(i)$		& $i$ & $a^N(i)$	& $i$ & $a^N(i)$ \\
	\midrule
	0 & (1\;1\;1\;1)	& 7 & (1\;3\;1\;3)	& 13 & (3\;1\;3\;1) \\
	1 & (1\;1\;1\;3)	& 8 & (1\;3\;3\;1)	& 14 & (3\;1\;3\;3) \\
	2 & (1\;1\;1\;5)	& 9 & (1\;3\;3\;3)	& 15 & (3\;3\;1\;1) \\
	3 & (1\;1\;3\;1)	& 10 & (1\;5\;1\;1)	& 16 & (3\;3\;1\;3) \\
	4 & (1\;1\;3\;3)	& 11 & (3\;1\;1\;1)	& 17 & (3\;3\;3\;1) \\
	5 & (1\;1\;5\;1)	& 12 & (3\;1\;1\;3)	& 18 & (5\;1\;1\;1) \\
	6 & (1\;3\;1\;1)\\
    \bottomrule
	\end{tabular}
	\label{tab::codebook-lut}
\end{table}

\subsection{Bounded Energy Trellis}

\begin{figure}
	\centering
    \input{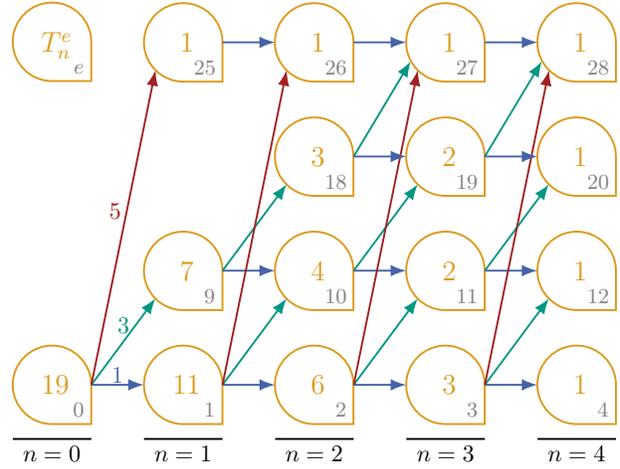}%
\newcount\b%
\begin{tikzpicture}[thick]
\tikzset{defaultnode/.style={draw,trellisnode,color=KITorange,inner sep=3pt, text width=2ex, align=right}}%
\matrix (trellis)
[matrix of nodes, nodes={trellisnode, draw}, column sep={0.75cm},
row sep={0.5cm}]
{%
\node[trellisnode, defaultnode](trellis-1-1){\nodepart{upper}{\large $T^e_n$}\nodepart[color=gray]{lower}\footnotesize$e$};
&\node[trellisnode, defaultnode](trellis-1-2){\nodepart{upper}{\large$1$}\nodepart[color=gray]{lower}\footnotesize$25$};
&\node[trellisnode, defaultnode](trellis-1-3){\nodepart{upper}{\large$1$}\nodepart[color=gray]{lower}\footnotesize$26$};
&\node[trellisnode, defaultnode](trellis-1-4){\nodepart{upper}{\large$1$}\nodepart[color=gray]{lower}\footnotesize$27$};
&\node[trellisnode, defaultnode](trellis-1-5){\nodepart{upper}{\large$1$}\nodepart[color=gray]{lower}\footnotesize$28$};\\
\node[coordinate](trellis-2-1){};
& \node[coordinate](trellis-2-2){};
&\node[trellisnode, defaultnode](trellis-2-3){\nodepart{upper}\large$3$\nodepart[color=gray]{lower}\footnotesize$18$};
&\node[trellisnode, defaultnode](trellis-2-4){\nodepart{upper}\large$2$\nodepart[color=gray]{lower}\footnotesize$19$};
&\node[trellisnode, defaultnode](trellis-2-5){\nodepart{upper}\large$1$\nodepart[color=gray]{lower}\footnotesize$20$};\\
\node[coordinate](trellis-3-1){};
&\node[trellisnode, defaultnode](trellis-3-2){\nodepart{upper}\large$7$\nodepart[color=gray]{lower}\footnotesize$9$};
&\node[trellisnode, defaultnode](trellis-3-3){\nodepart{upper}\large$4$\nodepart[color=gray]{lower}\footnotesize$10$};
&\node[trellisnode, defaultnode](trellis-3-4){\nodepart{upper}\large$2$\nodepart[color=gray]{lower}\footnotesize$11$};
&\node[trellisnode, defaultnode](trellis-3-5){\nodepart{upper}\large$1$\nodepart[color=gray]{lower}\footnotesize$12$};\\
\node[trellisnode, defaultnode](trellis-4-1){\nodepart{upper}\large$19$\nodepart[color=gray]{lower}\footnotesize$0$};
&\node[trellisnode, defaultnode](trellis-4-2){\nodepart{upper}\large$11$\nodepart[color=gray]{lower}\footnotesize$1$};
&\node[trellisnode, defaultnode](trellis-4-3){\nodepart{upper}\large$6$\nodepart[color=gray]{lower}\footnotesize$2$};
&\node[trellisnode, defaultnode](trellis-4-4){\nodepart{upper}\large$3$\nodepart[color=gray]{lower}\footnotesize$3$};
&\node[trellisnode, defaultnode](trellis-4-5){\nodepart{upper}\large$1$\nodepart[color=gray]{lower}\footnotesize$4$};\\
};%
\foreach[count=\a] \b in {2,...,5}{%

    \ifnum\a>1{
    \draw [KITblue, -{Latex}] (trellis-4-\a.east) -- (trellis-4-\b.west);%
    \draw [KITgreen, -{Latex}] (trellis-4-\a.east) -- (trellis-3-\b.south west);
    \draw [KITgreen, -{Latex}] (trellis-3-\a.east) -- (trellis-2-\b.south west);
    \draw [KITblue, -{Latex}] (trellis-3-\a.east) -- (trellis-3-\b.west);%
    \draw [KITblue, -{Latex}] (trellis-1-\a.east) -- (trellis-1-\b.west);%
    \draw [KITred, -{Latex}] (trellis-4-\a.east) -- (trellis-1-\b.south west);%
    \ifnum\a>2{%
    \draw [KITgreen, -{Latex}] (trellis-2-\a.east) -- ([xshift=-3,yshift=3.7]trellis-1-\b.south west);%
    \draw [KITblue, -{Latex}] (trellis-2-\a.east) -- (trellis-2-\b.west);%
    }\fi%
    }%
    \else{%
    \draw [KITblue, -{Latex}] (trellis-4-\a.east) -- node[above, yshift=-0.1cm]{$1$} (trellis-4-\b.west);%
    \draw [KITgreen, -{Latex}] (trellis-4-\a.east) -- node[midway, above]{$3$} (trellis-3-\b.south west);%
    \draw [KITred, -{Latex}] (trellis-4-\a.east) -- node[midway, above left, xshift=0.1cm]{$5$} (trellis-1-\b.south west);%
    }%
    \fi%
    \draw [black] ([yshift=-0.8cm]trellis-4-\a.west)--([yshift=-0.8cm]trellis-4-\a.east);%
}%
\node[rectangle, below= 0.2of trellis-4-1.south]{$n=0$};%
\node[rectangle, below= 0.2 of trellis-4-2.south]{$n=1$};%
\node[rectangle, below= 0.2 of trellis-4-3.south]{$n=2$};%
\node[rectangle, below= 0.2 of trellis-4-4.south]{$n=3$};%
\node[rectangle, below= 0.2 of trellis-4-5.south]{$n=4$};%
    \draw [black] ([yshift=-0.8cm]trellis-4-5.west)--([yshift=-0.8cm]trellis-4-5.east);%
\end{tikzpicture}
    \vspace*{-2ex}%
	\caption{Bounded energy trellis diagram for $N = 4$ and $E_\text{max} = 28$ and $8$-ASK following \cite{Gultekin2020}.}
	\label{fig::trellis}
\end{figure}

Storing the codebook in a \gls{lut}, as in the example in Table~\ref{tab::codebook-lut}, quickly becomes impractical for large codebooks.
However, it is not necessary to explicitly store the codebook; we  only require a fast way of finding how many sequences are lexicographically below a given sequence.
This can be achieved by a bounded energy trellis.
It consists of nodes corresponding to a number of amplitudes $n$ and accumulated energy $e$.
Each node $T_n^e$ holds the number of different sequences that are still possible with $n$ fixed amplitudes which result in the accumulated energy $e$.
For example, by using the same base parameters as for Table~\ref{tab::codebook-lut} the node $T_3^{19}$ has two possible continuations i.e. $T_3^{19} = 2$.
As $n = 3$ amplitudes are fixed, only $N - n = 1$ amplitude can be varied.
This amplitude could take the values $1$ or $3$.
However, if it takes the value $5$ or higher the total energy $e + 5^2 = 19 + 5^2 = 44$ would exceed the maximum energy $E_\text{max} = 28$.
Thus the number of possible continuations is two.
An example trellis for $N = 4$, $E_\text{max} = 28$ and $8$-ASK can be seen in Figure~\ref{fig::trellis}.
It holds the same codebook and results in the same mapping as Table~\ref{tab::codebook-lut}.
The number of accumulated energy values can be reduced by observing that the energy added by any amplitude can be written in the form $1 + k \cdot 8$.
For instance, the amplitude $5$ has the energy $5^2 = 25 = 1 + 3 \cdot 8$.
Thus the accumulated energies after $n$ fixed amplitudes will always be $n$ plus a multiple of $8$ and trellis nodes are only needed for these values.
Of course, there must be nodes for $n = 0$ to $n = N-1$ fixed amplitudes.
There are also nodes for $n = N$ fixed amplitudes.
Trivially, all of these nodes have the value $1$ and the remaining values in the trellis are built up backwards from these.
Assuming we have the values in all nodes with $n + 1$ fixed amplitudes, the value of a node with $n$ fixed amplitudes is the sum of all values from nodes with $n + 1$ fixed amplitudes which are reachable by adding a single amplitude to the node.
Adding an amplitude to a node means adding its energy to the accumulated energy of the node and corresponds to appending the amplitude to the sequences represented by this node.
This rule holds because the value of a node is the number of different sequences possible with it as the starting point.
Naturally, the number of continuations is the sum of the number of continuations after each possible next amplitude.
By leveraging the fact that the nodes with $n = N$ are indeed all known to be $1$, all values in the trellis can thus be calculated by applying
\begin{equation}
	\label{eq::trellis-construction}
	T_n^e = \sum_{a \in \mathcal{A}} T_{n+1}^{e+a^2}
\end{equation}
recursively starting from $n =N-1$ and down to $n = 0$.

\subsection{Encoding and Decoding via the Trellis Diagram}

Each amplitude sequence can be interpreted as a path through the bounded energy trellis: each transition in the trellis is equivalent to appending an amplitude to a sequence represented by a node.
For example, the amplitude sequence $(1 \; 3 \; 3 \; 1)$ corresponds to the path $T_0^0 \to T_1^1 \to T_2^{10} \to T_3^{19} \to T_4^{20}$ in the trellis diagram in Figure~\ref{fig::trellis}.
Indexing the sequences in the trellis makes use of this path representation and the definition of the index being the number of lexicographically lower sequences.
Amplitude sequences are constructed from left to right, therefore the sequences have their more significant amplitudes added first.

The index of a given sequence is defined by the lexicographical ordering but the sequence for a given index is only defined as the inverse operation.
Therefore, it is best to discuss the decoding algorithm (finding the index of a given sequence) first.
Indexing a sequences in the trellis makes use of its path representation by following the path one step at a time.
This corresponds to ``building'' the sequence by appending one amplitude in each step.
By keeping track of the number of sequences left lexicographically below in each step, the sum of lower sequences can be computed, which is the index.
The number of sequences left below in each step is the number of sequences possible if a lower amplitude would be appended instead of the next one in the sequence.
For each lower amplitude $a$, this number can easily be retrieved from the trellis diagram as it is the value of the trellis node reached if the lower amplitude $a$ is used next.
Thus if we are currently in the node $T_n^{e_n}$, the number of sequences possible if amplitude $a$ is appended equals $T_{n+1}^{e_n + a^2}$.
Algorithm~\ref{alg::seqtoidx} accumulates the number of possible sequences for each lower amplitude in each step to compute the index of a given sequence.
For the chosen system parameters in Figure~\ref{fig::trellis}, the obtained indices correspond to the codebook in Table~\ref{tab::codebook-lut}.

\begin{algorithm}
	\caption{Mapping Amplitude Sequence to Index}
	\label{alg::seqtoidx}
	\begin{algorithmic}
		\INPUT $\bm{a}^N$
		\STATE $e_n =
			\begin{cases}
				0,                      & n = 0\\
				\sum_{j=0}^{n-1} a_n^2, & n \in \{1, ..., N-1\}
			\end{cases}$
		\STATE $i = 0$
		\FOR{$n = 0$ to $N-1$}
			\FOR{$a \in \mathcal{A};\; a < a_n$}
			\STATE $i = i + T_{n+1}^{e_n + a^2}$
			\ENDFOR
		\ENDFOR
		\OUTPUT $i$
	\end{algorithmic}
\end{algorithm}

Encoding, which is mapping an index to an amplitude sequence can be achieved using Algorithm~\ref{alg::idxtoseq}.
If the path of a full length amplitude sequence contains a node $T_n^e$, it also contains one of the $T_n^e$ continuations of this node.
The sequence cannot be lexicographically greater than all its continuations after the amplitude in location $n$.
Thus the index of a sequence with node $T_n^e$ in its path is upper bounded by $T_n^e - 1$ plus the number of sequences left lexicographically below in the path leading up to node $T_n^e$.
Finding the correct next node now becomes a matter of finding the lowest next amplitude such that the value of the next node plus the number of lexicographically lower sequences is greater than the index.
For example, assume we are searching for the sequence belonging to index $i = 13$ using the trellis in Figure~\ref{fig::trellis}.
We know it starts with $(3 \; 1 \; ? \; ?)$ and $j = 11$ sequences are lexicographically lower than this start of the sequence.
Following the path or calculating the accumulated energy ($3^2 + 1^2 = 10$) shows that we are on a path that currently ends on node $T_2^{10}$.
If the next amplitude is chosen to be $1$, the next node is $T_3^{11}$.
This will lead to an index which is too small as $j + T_3^{11} = 11 + 2 = 13 \leq 13 = i$.
Thus, the next larger amplitude $3$ must be tried.
As all sequences continuing with $1$ are lexicographically below any sequence continuing with a $3$, these must be added to the number of sequences left below.
The variable $j$ is thus updated by adding $T_3^{11} = 2$, which is the number of sequences continuing with amplitude $1$.
The current index $j$ now equals $11 + 2 = 13$.
If the next amplitude is a $3$, the next node is $T_3^{19}$.
Now the index $i = 13$ is smaller than $j + T_3^{19} = 13 + 2 = 15$ and the next amplitude is chosen to be $3$.
Checking Table~\ref{tab::codebook-lut} shows that the correct sequence with index $i = 13$ is $(3 \; 1 \; 3 \; 1)$, which does indeed have a $3$ in the location in question.
Algorithm~\ref{alg::idxtoseq} starts from the known starting node $T_0^0$ and applies this method iteratively to compute the full sequence.

\begin{algorithm}
	\caption{Mapping Index to Amplitude Sequence}
	\label{alg::idxtoseq}
	\begin{algorithmic}
		\INPUT $i$
		\STATE $\bm{a}^N = (a_0 \; a_1 \; \cdots \; a_{N-1}) \in \mathcal{A}^N$
		\STATE $e = 0$
		\STATE $j = 0$
		\FOR{$n = 0$ to $N-1$}
			\STATE $a = 1$
			\WHILE{$i \geq j + T_{n+1}^{e + a^2}$}
				\STATE $j = j + T_{n+1}^{e + a^2}$
				\STATE $a =$ next larger value in $\mathcal{A}$
			\ENDWHILE
			\STATE $a_n = a$
			\STATE $e = e + a^2$
		\ENDFOR
		\OUTPUT $a^N$
	\end{algorithmic}
\end{algorithm}

\subsection{Amplitude Distribution and Average Energy}

For the purpose of evaluating the resulting code book, two metrics are especially interesting: the amplitude distribution and the average symbol energy.
The amplitude distribution is defined as the probability of finding a given amplitude in a random location in a sequence chosen randomly from the codebook.
As \gls{ess} is an indirect method and thus uses no predefined amplitude distribution, the amplitude distribution must be calculated from the codebook.
As shown in~\cite{Gultekin2020}, the amplitude distribution can be calculated using the trellis representation via

\begin{equation}
	\label{eq::amplitude-distribution}
	P_A(a) = \frac{T_1^{a^2}}{T_0^0}.
\end{equation}

The average energy can be computed by averaging the energy of amplitude sequences in the codebook.
It is of interest because it directly influences the signal-to-noise ratio in the case of an \gls{awgn} channel..
Given the amplitude distribution, the average energy
\begin{equation}
	\label{eq::average-energy}
	E_\text{av} = N \sum_{a \in \mathcal{A}} P_A(a) a^2
\end{equation}
can trivially be computed using the energies of the amplitudes and the sequence length~\cite{Gultekin2020}.

\section{Fixed-Length Messages and Optimum ESS} \label{sec::oess} %

In the general case, the number of sequences in the codebook is not a power of two.
This is disadvantageous as in a fixed-to-fixed distribution matcher, a fixed number of bits should be mapped to these sequences and the number of possible bit strings of any length is always a power of two.
Using the binary interpretation of the bit stream as index, sequences that have an index higher than $2^{N_\text{bit}} - 1$ are not used.
For large codebooks this disadvantage becomes negligible. For very small codebooks, however, \gls{ess} becomes less efficient than other methods.
As the sequences are ordered lexicographically and not by their total energy, the sequences with the highest indices do not necessarily have the highest energy.
This removes lower energy sequences from the codebook and the average energy is no longer minimal.
The rate loss incurred by \gls{ess} compared to an optimal minimum average energy codebook, is hereby increased.
\Gls{oess} as proposed in \cite{Chen2022} alleviates this problem.

As multiple energy thresholds $E_\text{max}$ can lead to the same possible bit string length, \gls{oess} is defined for the lowest $E_\text{max}$ that leads to a given bit string length.
The key idea of \gls{oess} is to use two trellis diagrams instead of only one.
One trellis diagram is a normal bounded energy trellis but with the threshold $E_\text{max} - 8$.
As discussed in the previous section, the energy of amplitude sequences is quantized to multiples of eight plus an offset.
Therefore, the first trellis diagram, called the full trellis in \cite{Chen2022}, contains all sequences except for those with maximum energy content $E_\text{max}$.
The second trellis diagram is called the partial trellis and contains only the sequences with energy equal to $E_\text{max}$.
A trellis like this can easily be constructed by altering the values of the final trellis nodes during initialization.
For a regular \gls{ess} trellis,  all nodes with $n = N$ are initialized to $1$.
In the partial trellis used by \gls{oess}, only $T_N^{E_\text{max}}$ is set to $1$ while all other nodes are set to $0$.
Applying \eqref{eq::trellis-construction} in the regular way calculates all other node values.
By splitting the sequences with maximum energy into a new trellis, these are enumerated separately.
Mapping an index to a sequence now works by first selecting the appropriate trellis.
If the index is lower than the number of sequences in the full trellis, a sequence from the full trellis is chosen using Algorithm~\ref{alg::idxtoseq}.
Otherwise a local index for the partial trellis is created by subtracting the number of sequences in the full trellis from the index.
Algorithm~\ref{alg::idxtoseq} is then used on the partial trellis with the local index to find the corresponding amplitude sequence with energy $E_\text{max}$.
This way the highest indices correspond to sequences with maximum energy.
Therefore, removing sequences which are located in the partial trellis will reduce the average energy.
De-mapping works in a similar way as the mapping.
First, the energy of the sequence is calculated.
If it is below the maximum energy, the full trellis is used with Algorithm~\ref{alg::seqtoidx} to find the index.
Should the sequence have maximum energy, the local index is calculated from the partial trellis using Algorithm~\ref{alg::seqtoidx} and the number of sequences in the full trellis is added to it, which is necessary to obtain the final index from the local index computed with the partial trellis.

To calculate the amplitude distribution in \gls{oess}, we are not able to use the simple form \eqref{eq::amplitude-distribution}. Instead, we need to apply a calculation that takes into account that some of the sequences are removed from the codebook, therefore changing the amplitude distribution. Calculation of the amplitude distribution for \gls{ess} with a limited codebook size and calculation of the amplitude distribution for \gls{oess} can be  found in \cite{Chen2022}.

\section{Introducing RSESS} \label{sec::rsess} %

Our contribution is a free and open source implementation of the \gls{ess} and \gls{oess} algorithms.
We used the programming language Rust to implement the presented algorithms. We distribute the code in the form of a Rust crate named RSESS on crates.io.
The full source code is also available at \url{https://github.com/kit-cel/rsess}.
Encoding and decoding between indices and amplitude sequences form the core of RSESS.
For analysis, calculation of the amplitude distribution is implemented both for the simple case in which all sequences are used as well as for the more complex cases in which only indexes up to a power of two are used or the amplitude distribution for \gls{oess}.
Calculation of the average energy is also implemented as well as the calculation of the energy distribution, which gives the probabilities for sequences of specific energy.
The programming interfaces to work with \gls{ess} or \gls{oess} are identical.
However, the \gls{oess} implementation features an additional function which can find the $E_\text{max}$ values for which \gls{oess} is defined.
To facilitate the use in Python scripts, Python bindings for RSESS are provided in a package named PyRSESS.
The source code resides in the same repository, but PyRSESS is also published to PyPI.
The Python bindings cover the full scope of the Rust library.

Using a Rust or Python package manager, either RSESS or PyRSESS is easy to install.
In both programming languages, we expose an object-oriented interface with one class for \gls{ess} and \gls{oess} each.
Objects instantiated from these classes can be used to encode and decode bit strings into amplitude sequences and vice-versa.
While RSESS uses the arbitrarily sized integers from the rug Rust crate as indices, the Python bindings use arrays/lists containing zeros and ones to model data bits.
Usage examples for both RSESS and PyRSESS are also made available at \url{https://github.com/kit-cel/rsess_examples}.

\begin{figure}
  \begin{tikzpicture}
    \tikzset{
      rsess/.style={solid},
      pyrsess/.style={dashed},
      ess/.style={KITblue, mark=square*, mark options={solid}},
      oess/.style={KITred, mark=triangle, mark options={solid}},
    }
  \begin{axis}[
    xlabel={Sequence length},
    ylabel={Run time (\si{ms})},
    ymin=50,
    ymax=20000,
    xmin=50,
    xtick={50, 1000, 2000, 3000},
    xmax=3200,
    ymode=log,
    grid=both,
    width=1\columnwidth,
    height=0.7\columnwidth,
    mark size=1.5pt,
    legend style={at={(axis cs:3150,50)},anchor=south east,nodes={transform shape}, font=\footnotesize},
    legend cell align={left},
    y tick label style={
      /pgf/number format/.cd,
      fixed,
      precision=1,
      /tikz/.cd
    },%
    ]%
    \addplot+[%
    line width=1.5pt,%
    ess, rsess%
    ]%
    table[discard if not={class}{ESS}, discard if not={tool}{RSESS},x=n_max, y=encoding_time_ms, col sep=comma,
    ]{./data/measurements-insl49_Intel_Core_i7-7700.csv};%
    \addplot+[%
    line width=1.5pt,%
    oess, rsess,%
    ]%
    table[discard if not={class}{OESS}, discard if not={tool}{RSESS}, x=n_max, y=encoding_time_ms, col sep=comma,
    ]{./data/measurements-insl49_Intel_Core_i7-7700.csv};
    \addplot+[%
    line width=1.5pt,%
    ess, pyrsess%
    ]%
    table[discard if not={class}{ESS}, discard if not={tool}{PyRSESS},x=n_max, y=encoding_time_ms, col sep=comma,
    ]{./data/measurements-insl49_Intel_Core_i7-7700.csv};
    \addplot+[%
    line width=1.5pt,%
    oess, pyrsess,%
    ]%
    table[discard if not={class}{OESS}, discard if not={tool}{PyRSESS}, x=n_max, y=encoding_time_ms, col sep=comma,
    ]{./data/measurements-insl49_Intel_Core_i7-7700.csv};
    \legend{ESS RSESS, OESS RSESS, ESS PyRSESS, OESS PyRSESS};
    \end{axis}%
    \end{tikzpicture}%
    \vspace*{-2ex}%
    \caption{Encoding times over varying sequences length for 10000 sequences using our framework}%
    \label{fig::encoding_time}%
    \vspace*{-2ex}%
\end{figure}
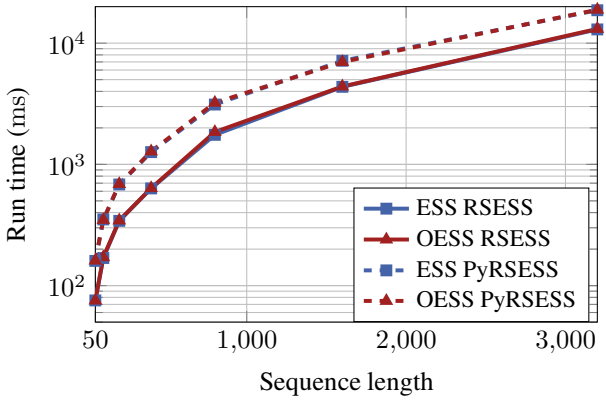
\begin{figure}
  \begin{tikzpicture}
    \tikzset{
      rsess/.style={solid},
      pyrsess/.style={dashed},
      ess/.style={KITblue, mark=square*, mark options={solid}},
      oess/.style={KITred, mark=triangle, mark options={solid}},
    }
  \begin{axis}[
    xlabel={Sequence length},
    ylabel={Run time (\si{ms})},
    ymin=10,
    ymax=20000,
    xmin=50,
    xtick={50, 1000, 2000, 3000},
    xmax=3200,
    ymode=log,
    grid=both,
    width=1\columnwidth,
    height=0.7\columnwidth,
    mark size=1.5pt,
    legend style={at={(axis cs:3150,50)},anchor=south east,nodes={transform shape}, font=\footnotesize},
    legend cell align={left},
    y tick label style={
      /pgf/number format/.cd,
      fixed,
      precision=1,
      /tikz/.cd
    },%
    ]%
    \addplot+[%
    line width=1.5pt,%
    ess, rsess%
    ]%
    table[discard if not={class}{ESS}, discard if not={tool}{RSESS},x=n_max, y=decoding_time_ms, col sep=comma,
    ]{./data/measurements-insl49_Intel_Core_i7-7700.csv};%
    \addplot+[%
    line width=1.5pt,%
    oess, rsess,%
    ]%
    table[discard if not={class}{OESS}, discard if not={tool}{RSESS}, x=n_max, y=decoding_time_ms, col sep=comma,
    ]{./data/measurements-insl49_Intel_Core_i7-7700.csv};
    \addplot+[%
    line width=1.5pt,%
    ess, pyrsess%
    ]%
    table[discard if not={class}{ESS}, discard if not={tool}{PyRSESS},x=n_max, y=decoding_time_ms, col sep=comma,
    ]{./data/measurements-insl49_Intel_Core_i7-7700.csv};
    \addplot+[%
    line width=1.5pt,%
    oess, pyrsess,%
    ]%
    table[discard if not={class}{OESS}, discard if not={tool}{PyRSESS}, x=n_max, y=decoding_time_ms, col sep=comma,
    ]{./data/measurements-insl49_Intel_Core_i7-7700.csv};
    \legend{ESS RSESS, OESS RSESS, ESS PyRSESS, OESS PyRSESS};
    \end{axis}%
    \end{tikzpicture}%
    \vspace*{-2ex}%
    \label{fig::decoding_time}%
    \caption{Decoding times over varying sequence length for 10000 sequences using our framework}%
    \vspace*{-2ex}%
\end{figure}
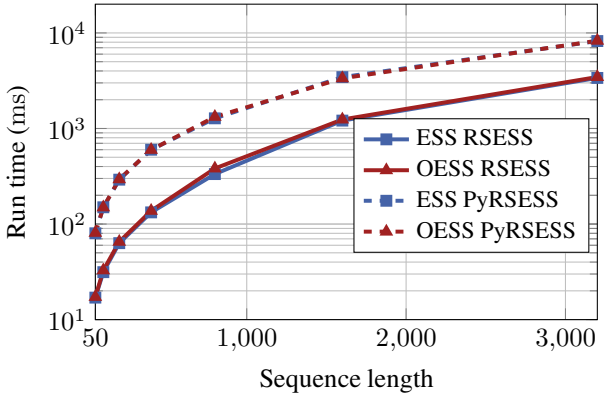

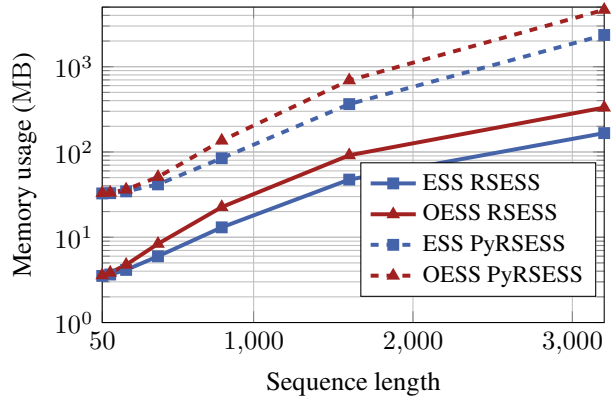
\begin{figure}
  \begin{tikzpicture}
    \tikzset{
      rsess/.style={solid},
      pyrsess/.style={dashed},
      ess/.style={KITblue, mark=square*, mark options={solid}},
      oess/.style={KITred, mark=triangle, mark options={solid}},
    }
  \begin{axis}[
    xlabel={Sequence length},
    ylabel={Memory usage (\si{MB})},
    ymin=1,
    ymax=5000,
    xmin=50,
    xtick={50, 1000, 2000, 3000},
    xmax=3200,
    ymode=log,
    grid=both,
    width=1\columnwidth,
    height=0.7\columnwidth,
    mark size=1.5pt,
    legend style={at={(axis cs:3150,2)},anchor=south east,nodes={transform shape}, font=\footnotesize},
    legend cell align={left},
    y tick label style={
      /pgf/number format/.cd,
      fixed,
      precision=1,
      /tikz/.cd
    },%
    ]%
    \addplot+[%
    line width=1.5pt,%
    ess, rsess%
    ]%
    table[discard if not={class}{ESS}, discard if not={tool}{RSESS},x=n_max, y=resident_MB, col sep=comma,
    ]{./data/measurements-insl49_Intel_Core_i7-7700.csv};%
    \addplot+[%
    line width=1.5pt,%
    oess, rsess,%
    ]%
    table[discard if not={class}{OESS}, discard if not={tool}{RSESS}, x=n_max, y=resident_MB, col sep=comma,
    ]{./data/measurements-insl49_Intel_Core_i7-7700.csv};
    \addplot+[%
    line width=1.5pt,%
    ess, pyrsess%
    ]%
    table[discard if not={class}{ESS}, discard if not={tool}{PyRSESS},x=n_max, y=resident_MB, col sep=comma,
    ]{./data/measurements-insl49_Intel_Core_i7-7700.csv};
    \addplot+[%
    line width=1.5pt,%
    oess, pyrsess,%
    ]%
    table[discard if not={class}{OESS}, discard if not={tool}{PyRSESS}, x=n_max, y=resident_MB, col sep=comma,
    ]{./data/measurements-insl49_Intel_Core_i7-7700.csv};
    \legend{ESS RSESS, OESS RSESS, ESS PyRSESS, OESS PyRSESS};
    \end{axis}%
    \end{tikzpicture}%
    \vspace*{-2ex}%
    \caption{Resident memory over varying sequence lengths for 10000 sequences using our framework}%
    \label{fig::memory}%
    \vspace*{-2ex}%
\end{figure}
The main reason for implementing the \gls{ess} algorithm in the compiled language Rust was the goal to have fast encoding and decoding.
Simulations regularly calculate thousands of transmissions over the simulated channels and
a fast implementation is invaluable in this situation.
In ESS encoding/decoding, speed mainly depends on the amplitude sequence length $N$, the energy threshold $E_\text{max}$, and the data itself.
Together, the combination of $N$ and $E_\text{max}$ determines the number of data bits $N_\text{bits}$.
For the following benchmarks a constant shaping rate $r_\text{sh} = \frac{N}{N_\text{bits}} = 1.5$ and the minimum $E_\text{max}$ possible with this $r_\text{sh}$ is used.
The use of $10000$ random data sequences allows statements about the data-independent average encoding / decoding behavior.
This allows the amplitude sequence length $N$ to be the only parameter influencing algorithm complexity.
Figure~\ref{fig::encoding_time} shows the duration of encoding $10000$ random bit strings for different values of $N$ and Figure~\ref{fig::decoding_time} shows the duration of decoding the resulting amplitude sequences back into bit strings.
As the main advantage of the ESS / OESS algorithm is its low rate loss for short block lengths, performance for short amplitude sequence lengths is especially relevant.
Decoding times for $10000$ sequences are below one second, even for very long sequence lengths up to $N \approx 1300$.
Using the Python bindings PyRSESS, decoding times stay below one second for sequence lengths up to $N \approx 600$.
Encoding long sequences with lengths of $N \approx 500$ is slower but still below one second.
The Python bindings only keep encoding below one second for medium block lengths below $N \approx 300$.
In general, decoding is faster than encoding and pure Rust is faster than using the Python bindings.
Another limiting factor may be the memory space used to store the trellis.
However, our benchmarks in Figure~\ref{fig::memory} show that using pure Rust, this is not the case as the total resident process memory does not exceed \SI{100}{\mega\byte} even for long sequences up to $N = 1600$.
All memory measurements were done directly after creating the trellis and captured the resident memory of the whole process, not only the trellis.
Unlike the encoding/decoding times, the memory usage values of \gls{ess} and \gls{oess} differ.
This is to be expected as \gls{oess} uses two trellises while \gls{ess} only uses one.
The memory usage of PyRSESS is much higher than that of the pure Rust library and even exceeds \SI{2}{\giga\byte} for \gls{ess} with the extreme block length of $N = 3200$.
\Gls{oess} has even higher memory usage exceeding \SI{4}{\giga\byte}, however the advantage of \gls{oess} over \gls{ess} already vanishes for far lower block lengths.
We would advise against the use of PyRSESS for simulations with extremely long block lengths on hardware with limited memory resources.

\begin{figure}
  \begin{tikzpicture}
    \tikzset{
      n40/.style={solid},
      n20/.style={dashed},
      ess/.style={KITblue, mark=square*, mark options={solid}},
      oess/.style={KITred, mark=triangle, mark options={solid}},
      uniform/.style={KITorange, mark=diamond, mark options={solid}},
      capacity/.style={black, dotted, mark=},
    }
  \begin{axis}[
    xlabel={$E_\text{s}/N_0$ (dB)},
    ylabel={$\text{AIR}_N$ (bit/symbol)},
    ymin=3.6,
    ymax=4.4,
    xmin=11.8,
    xtick={11.5, 12, 12.5, 13, 13.5, 14, 14.5},
    xmax=14.7,
    grid=both,
    width=1\columnwidth,
    height=0.7\columnwidth,
    mark size=1.5pt,
    legend style={at={(axis cs:14.6,3.65)},anchor=south east,nodes={transform shape}, font=\footnotesize},
    legend cell align={left},
    y tick label style={
      /pgf/number format/.cd,
      fixed,
      precision=1,
      /tikz/.cd
    },%
    ]%
    \addplot+[%
    line width=1.5pt,%
    ess, n20%
    ]%
    table[x=E_s/N_0 in dB, y=ESS N20, col sep=comma,
    ]{./data/comparison_to_chen2022.csv};%
    \addplot+[%
    line width=1.5pt,%
    ess, n40%
    ]%
    table[x=E_s/N_0 in dB, y=ESS N40, col sep=comma,
    ]{./data/comparison_to_chen2022.csv};%
    \addplot+[%
    line width=1.5pt,%
    oess, n20%
    ]%
    table[x=E_s/N_0 in dB, y=OESS N20, col sep=comma,
    ]{./data/comparison_to_chen2022.csv};%
    \addplot+[%
    line width=1.5pt,%
    oess, n40%
    ]%
    table[x=E_s/N_0 in dB, y=OESS N40, col sep=comma,
    ]{./data/comparison_to_chen2022.csv};%
    \addplot+[%
    line width=1.5pt,%
    uniform%
    ]%
    table[x=E_s/N_0 in dB, y=Uniform, col sep=comma,
    ]{./data/comparison_to_chen2022.csv};%
    \addplot+[%
    line width=1.5pt,%
    capacity%
    ]%
    table[x=E_s/N_0 in dB, y=Capacity, col sep=comma,
    ]{./data/comparison_to_chen2022.csv};%
    \legend{ESS $N=20$, ESS $N=40$, OESS $N=20$, OESS $N=40$, Uniform, Capacity};
    \end{axis}%
    \end{tikzpicture}%
    \vspace*{-2ex}%
    \caption{AIR over varying $E_\mathrm{s}/N_0$ (reproducing Figure 16 in~\cite{Chen2022})}%
    \label{fig::comparison_chen}%
        \vspace*{-2ex}%
\end{figure}
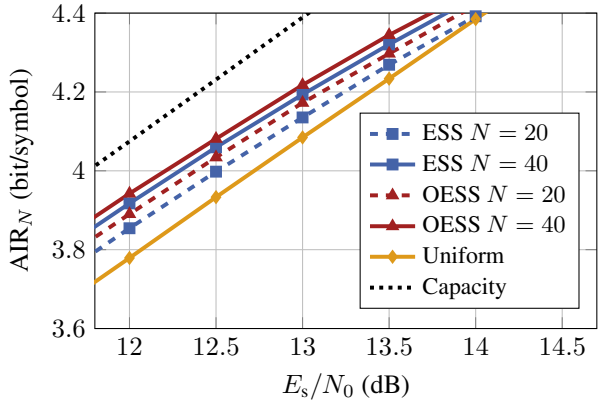

Multiple simulations over an \gls{awgn} channel were conducted using PyRSESS with different sequence lengths and energy thresholds.
Most notably, one simulation aimed at validating the \gls{air} results for \gls{ess}, \gls{oess}, and \gls{ccdm} at different signal-to-noise ratios published in~\cite{Chen2022}.
The \gls{air} is the maximum information rate that can reliably be transmitted over a channel assuming optimal channel coding and can be estimated from the soft information before channel decoding. %
Using PyRSESS for the simulation of \gls{ess} and \gls{oess}, we could replicate the results published in Figure~16 in~\cite{Chen2022}.
Our results can be seen in Figure~\ref{fig::comparison_chen}. %
Apart from validating the research by Yizhao Chen and colleagues, this also demonstrates that our implementations of the \gls{ess} and \gls{oess} algorithms are correct.

\vspace*{-1.5ex}
\section{Conclusion} \label{sec::conclusion} %
\balance
We have provided a short overview of probabilistic shaping and the \gls{ess} algorithm to then introduce our contribution: a free and open source implementation of \gls{ess} and \gls{oess} called RSESS.
RSESS is a Rust library and also has Python bindings called PyRSESS.
We have shown that RSESS is fast and memory efficient even for large simulations, while PyRSESS is an easy-to-use option for normal simulations but is less efficient and becomes demanding for very large simulations.
Finally, the functionality of our implementation could be verified by replicating literature results in the short block length regime.
This makes RSESS a viable tool for research and development in the field of probabilistic constellation shaping.

\section*{Acknowledgment}
This work has received funding from the European Research Council (ERC)
under the European Union's Horizon 2020 research and innovation programme (grant agreement No. 101001899).

\makeatletter
\interlinepenalty=10000

\bibliography{example_paper}
\bibliographystyle{grcon}
\makeatother
\end{document}